\documentstyle[11pt]{article}
\normalbaselines
\begin{document}

\begin{center}

{\LARGE\bf Statistical Topology of Real Polynomials.\\[0.5cm] 
I: Two Variables}

\vskip 1cm

{\large {\bf George Khimshiashvili}}\ and
{\large {\bf Alexander Ushveridze}}\footnote{This work was
partially supported by Deutsche Forschungsgemeinschaft under
grant N }

\vskip 0.1 cm

Department of Theoretical Physics, University of Lodz,\\
Pomorska 149/153, 90-236 Lodz, Poland\footnote{E-mail
addresses: khimiash@mvii.uni.lodz.pl and alexush@krysia.uni.lodz.pl}\\

\end{center}
\vspace{1 cm}
\begin{abstract}

We investigate average gradient degree of normal random polynomials of 
fixed algebraic degree $n$. In particular, for polynomials of two variables, 
asymptotics of the average gradient degree for large values of $n$ is
determined. 

\end{abstract}

\newpage

\section{Introduction}

Since the seminal work of M.Kac \cite{kac} random polynomials became an object
of permanent study in mathematics and physics \cite{bhr}, \cite{bog},
\cite{spe}. For polynomials of one variable, several important numerical
characteristics,
including average real roots number \cite{edk}, are already understood well
enough.

Much less is known about random polynomials of several variables. In fact,
only recently an important breakthrough was made by M.Shub and S.Smale 
\cite{ssm}, who have found a suitable probabilistic setting in higher
dimensions and, in particular, computed the average real roots number
having thus generalized the aforementioned classical results of M.Kac. Their 
results were, in turn, clarified and developed by E.Kostlan \cite{kos}.
The state of arts field is described in a review paper of A.Edelman
and E.Kostlan \cite{edk}.

It should be emphasized that despite these important contributions, many
natural problems for random polynomials of several variables still remain
completely uninvestigated. Up to the authors' knowledge, there are no
investigations of numerical invariants specific for higher dimensions,
such as Betti numbers of level surfaces or topological mapping degree.
At the same time, random polynomials inevitably appear in various mathematical
and physical problems of intrinsically multidimensional nature, where
consideration of such characteristics becomes urgent. This direction
of research may be (and in our exposition actually will be) called
{\em statistical topology of random polynomials}, in accordance with the title 
of this text. As usual, an important aspect of such investigations
is concerned with asymptotical behaviour of statistical characteristics.

In the present note we are mainly interested in {\em gradient degree} 
(topological degree of gradient mapping \cite{avg}) which is perhaps the 
simplest and most visual topological characteristic of multidimensional 
polynomials. As is well known, gradient degree plays an important role in 
the qualitive study of polynomial topology. In particular, it enables one 
to compute certain topological invariants of level surfaces. For example, 
from results of 
\cite{gkh} it follows readily that knowledge of the average of gradient degree 
enables one to compute 
the average Euler characteristic of level surfaces of random polynomials.

In other words, the average gradient degree definitely belongs
to the core of (the just christened discipline of) statistical
topology and it seemed appropriate for us to start with its investigation 
in the simplest case of two variables. 

Our concrete aim was to investigate the asymptotics of {\em average of modulus 
of gradient
degree} (which we will call simply {\em average absolute degree}) for 
certain natural distributions of coefficients of a random real polynomial in
two variables and 
below we describe some concrete results in this direction.

Actually, we work in a wider context of computing the
{\em average absolute topological degree} of a random polynomial mapping
defined by a pair of random polynomials
with independent identically distributed (i.i.d.) coefficients.
Since topological degree is nothing else than the
algebraic number of real solutions (counted with weights equal to the 
signs of Jacobian determinant) of the corresponding system of algebraic
equations, our results may be considered as complementary to those about 
the average real
roots number obtained by S.Shub, S.Smale and E.Kostlan. In this note we
only present first results in this direction, leaving their development and
applications for further publications. Also, no attempt is made to provide
formally exhaustive proofs because we are going to present full proofs
of more general results in the forthcoming paper. As a rule, we only outline
methods and indicate crucial points of proofs, to the extent hopefully
sufficient for making it clear why those results are true. 
Although the whole setting was suggested by physical motivations those will 
not be discussed here.

Results of this paper were obtained during the period when both authors were
Visiting Professors at the Department of Theoretical Physics of the University
of Lodz. We gratefully acknowledge hospitality and support of the whole staff
of the Department.

\section{General setting}

We will now describe a method of computing $E(\mid deg\ grad \mid)$ for most 
natural distributions of coefficients. Having in view physical applications we
mainly deal with the case when coefficients are independent standard normal
random variables. For small values of algebraic degree of polynomials
we will compute $E(\mid deg\ grad \mid)$ explicitly. We also reformulate
related probabilistic problems in terms of random walks and formulate a
general
hypothesis on asymptotics of the average gradient degree for large values
of algebraic degree of random polynomials under consideration.

We start by considering polynomials in two indeterminates. Thus let
$F_n$ be the ring of real polynomials in two indeterminates of the algebraic
degree $n$. Every $P\in F_n$ may be written in the standard form
$P(x,y) = \sum_{k+l=0}^n a_{kl} x^k y^l$ with $k,l \in \bf Z_+$. As usual, 
the gradient of such polynomial is a couple of polynomials
$grad\ P = (\partial P/\partial x,  \partial P/\partial y)$. Sometimes
we will denote it by $P'$ and identify with the corresponding polynomial 
endomorphism of $\bf R^2$. 

Recall that if $grad\ P$ is a proper mapping, i.e. full
preimage of every compact set is compact, then the topological degree
$deg\ grad\ P$ is well defined as the algebraic number of preimages of
any regular value of the gradient mapping \cite{miw}. As is well known,
the property of being proper is generic. In other words, polynomials with
non-proper gradients are "rare" and can be ignored in any probabilistic
considerations on $F_n$. More precisely, such "bad" polynomials constitute
an algebraic subset, hence a subset of measure zero in $F_n$ \cite{avg}.

Consider now an $F_n$-valued random variable $P(\omega)$ with $a_{kl}(\omega)$
certain real valued random variables. We will always assume that those
random variables are (stochastically) independent \cite{fe} (e.g. i.i.d.
coefficients). Often $a_{kl}(\omega)$ will be taken to be central normals 
(Gaussian random variables with zero mean and given covariance matrix
\cite{fe}) 
and then $P(\omega)$ is called simply a normal random polynomial.

For almost all values of $\omega$ one obtains a proper gradient mapping with the 
well defined topological degree and we are interested in statistical properties 
of the corresponding random variable $gd_P(\omega) = deg\ grad\ P(\omega)$.  
Of course, its average vanishes for trivial symmetry reasons but it is the 
average of its modulus which is really important in applications. Thus
we denote the latter by $E(agd_P)$ or $\widetilde{agd}_P$ and try to compute 
it explicitly, at least for some specially chosen distributions of coefficients. 
Then it becomes possible to determine its asymptotical behaviour and there is
some hope that the latter remains valid for a wider variety of situations,
as often happens with asymptotical results. 

It turns out that we can perform this programme even in a more general context 
of average topological degree $E(deg f(\omega))$ for certain normal random 
polynomial endomorphisms of $\bf R^2$
defined by a pair of i.i.d. random polynomials $(P(\omega),Q(\omega))$,
and this is what we are going to describe in the sequel. Results 
about $\widetilde{agd}$ will appear as corollaries.

\section{Useful reductions}

We want to simplify our task by reducing this problem to consideration
of {\em random binary forms}, that is random homogeneous polynomials in two
indeterminates with independent distributions of coefficients.
The possibility of such reduction follows from a simple topological
lemma. Recall
that the {\em leader} $P^*$ of a polynomial $P\in F_n$ is defined to be 
the binary form equal to the sum of monomials of the highest degree actually
entering into $P$. Of course, its degree may be less than $n$. Recall that
a polynomial endomorphism $(p,Q)$ of the plane is named non-degenerate,
if the two leaders $P^*, Q^*$ have no non-trivial common zeroes. It is
evident that non-degeneracy is a "generic" property so in computations
of averages one can deal only with non-degenerate pairs $(P,Q)$.

{\bf Lemma 1}. If $f=(P,Q)$ is a non-degenerate polynomial endomorphism of 
the plane then $deg f = deg f^*$, where $f^* = (P^*,Q^*)$.

The proof is very simple. First of all, a non-degenerate mapping is proper,
which follows from the well-known fact that the sum of moduli of leaders 
of a non-degenerate endomorphism majorates terms of lower degree on circles
of large radii. Thus the degree is well defined. Moreover, the conclusion 
follows from the same fact by application of Rouch\'{e}t principle.

As was noticed, while computing $E(deg)$ we can exclude degenerate mappings
and then the degree is always equal to that of the leader. Also, the set
of those polynomials with leaders having not the maximal possible degree 
has the measure zero, so we can consider only pairs of polynomials with 
leaders both having the degree $n$. Thus we see that it is sufficient 
to solve the problem for random binary forms. In fact, all what was said
refers to any reasonable distributions of coefficients. In further analysis
we restrict ourselves with three types of independent continuous distributions 
of coefficients, which already appeared in previous works on random polynomials:

1) standard normals \cite{fe};

2) central normals with special covariance matrix from \cite{ssm};

3) i.i.d. coefficients uniformly distributed on (-1,1).

All of them have common features with respect to our issue and results 
on asymptotics of average degree should be similar. 
Thus we will mainly deal with the case of 
independent standard normals and briefly mention necessary changes in other 
cases. Similar results may be also obtained for various discrete distributions 
of coefficients, where the situation is even simpler. In fact, for certain 
simplest discrete central distributions, such as symmetric $\{-1,1\}$-valued 
coin, the problem becomes purely combinatorial and for small values of $n$
may be solved explicitly with the aid of computer but we will not dwell on 
these elementary versions.

From now on we concentrate on binary forms and perform further
simplifications. It turns out helpful to introduce a natural subdivision 
of the space of events according to the number of real zeroes of binary 
forms considered. 

Again, by "genericity" arguments it is clear that our forms
can be supposed to have only simple zeroes. Evidently, the number of these
zeroes can be an arbitrary integer $r$ not exceeding $n$ and having the same
parity as $n$. We denote by $A_{rs}$ such subset of events when the first 
leader has precisely $r$ real zeroes and the second one has $s$ real zeroes.

From the very definition of the average it is clear that it is equal
to the weighted sum of averages over subsets $A_{rs}$ with the weights
equal to probabilities of subsets $A_{rs}$. Thus our task may be performed
in two steps:

1. compute probabilities of events $A_{rs}$;

2. compute the average $E(deg)$ with the respect to subset $A_{rs}$.

The first step is a purely geometrical problem and reduces to estimation
of certain integrals over multidimensional domains appearing as components
of complement to the discriminant hypersurface \cite{avg}. For small values 
of $n$ these computations may be done explicitly. In general this is 
a difficult analytical problem \cite{ssm}. Luckily, it turns out that precise 
values of these probabilities are not important in studying asymptotics 
of the average degree, as will be explained below.

Thus, it is the second step which we develop here in some detail. Our main 
tool is combinatorics arising from a natural analogy of the problem
with random walks on the real line. The latter issue is well studied
and necessary combinatorical results together with asymptotical analysis 
may be successfully borrowed from the theory of random walks. We benefit 
from this circumstance in the next section.

Before proceeding with combinatorics we would like to point out that
appearance of one-dimensional random walks in our problem is not occasional 
and may be illustrated by the following heuristical argument. Recent results 
of \cite{ssm} suggest that in certain cases the most probable number 
of common real zeroes of the random pair $(P,Q)$ is $n$. This suggests that 
one could restrict himself with only such pairs and try to estimate
average degree over this "representative" subset of events. In doing so
it is reasonable to suggest that at every root the probability of any of
two possible values of the sign of Jacobian $J(P,Q)$ is one half.

Of course, these probabilistical assumptions should be justified but accepting 
them for a moment, it becomes evident that we deal with the estimation of 
average deviation of paths in a symmetric random walk on the line and one can
use known results \cite{fe}. We found it remarkable that asymptotics obtained 
in this way is an agreement with precise results of the next section, which 
supports validity of such "oversimplified" approach. 

The big advantage of this approach is that the same heuristical considerations 
are applicable in all dimensions. In particular, inspired by asymptotical 
consistence of the both methods in two-dimensional case, we were led to certain 
general hypotheses on asymptotics of averages. One of such hypotheses
is presented in the conclusion of the paper.

\section{Average degree via random walks}

First we reduce computation of topological degree for a non-degenerate pair 
of binary forms to a simple geometric procedure. Recall that due to 
homogeneuity, zero sets of binary forms consist of straight lines containing 
the origin. Paint all zero-lines of the first binary form $P$ with one colour,
and choose a different colour for zero-lines of the second form $Q$. Then
we can orient the unit circle $T\subset \bf C$ and walk around it in the chosen
direction counting at the same time the number $d$ of alternances of colours 
inductively, as follows. 

If after a line of one colour comes a line of the other colour, and
only in this case, we increase the value of $d$ by one. In other words, 
pairs of consecutive lines of the same colour do not contribute to $d$.
It is trivial to check that the resulting value of $d$ does not depent on
the line from which to start. Thus with any pair of binary forms we may
associate a natural number.

{\bf Lemma 2}.  If the pair $(P,Q,)$ is non-degenerate then 
$\mid deg (P,Q)\mid = d$.

The proof again follows from invariance properties of topological degree.
First of all, notice that any two consecutive lines of the same colour
may be brought into coincidence by a non-degenerate homotopy creating no
common zeroes. The degree is unchanged under this procedure and now we 
have a square of linear form in the representation of our binary form.
As non-negative factors of components do not affect the value of degree,
we can omit these two lines and proceed inductively. It is clear that
eventually we reach the situation when numbers of lines of each colour
coincide, say are equal to certain $k$, and colours are mutually alternating
along unit circle. In order to compute degree in this case it remains 
to notice that such configuration of lines is non-degenerately homotopic
to zero-lines configuration of "realification" of the map 
$z^k:{\bf C} \rightarrow {\bf C}, z \mapsto z^k$.  The latter map evidently 
has the degree equal to $k$, which finishes the proof.

This understood, we come very close to the desired reduction to a 
one-dimensional
problem. Namely, we dehomogenize our forms by dividing over one of variables
and immediately observe that computation of $d$ is equivalent to computation
of average alternance number of real roots of two i.i.d. random polynomials of 
$n$-th degree. In order to avoid troubles with probabilities of $A_{rs}$
let us consider first {\em hyperbolic} polynomials, that is those having all
their roots situated on the real axis. An equivalent way of saying this is that 
we consider
$n$-forms represented as products of linear forms with i.i.d coefficients. 
For distributions listed in previous section coefficients of these linear forms 
are centered (have vanishing means) and it turns out that their explicit form 
is not important.

Indeed, it becomes clear that the symmetrical role of both polynomials implies  
that the number of alternances under consideration may be interpreted in terms 
of a symmetric random walk on the real line. Dealing with random walks we will 
freely use standard geometric terminology of \cite{fe}. 

Thus we consider a symmetric random walk on the real line and want to compute 
the average number of alternances, in the sense explicated above, over the set 
of all trajectories of length $2n$ ending on the horizontal axis. This may
be done explicitly via simple combinatorics. Namely, the number of such 
trajectories is evidently equal to 

\[
C^{2n}_n = \frac {(2n)!}{n!n!} .
\]

Denoting by $\epsilon_k$ discrete entities equal to $1$ or $-1$ we see
that all what is needed is to compute the sum
\[
\sum (-1)^k \epsilon_k
\]
with the condition that sum of all $\epsilon_k$ is equal to zero.

{\bf Lemma 3}. The average value of alternances is equal to
\[
\frac {\sum_{k=0}^n  (C^{n}_{k})^2 \cdot \mid 2k - n \mid}{C^{2n}_n} .
\]

In order to obtain this formula, observe first that the value of the above
sum is evidently determined by the number $k$ of $\epsilon_j=1$ with
odd $j$. In fact, then the degree is found to be equal to $\mid 2k - n \mid$. 
Now, these $k$ places are arbitrary within $n$ odd numbers from $1$ to $2n-1$,
and the same holds for the choice of places of the rest $n-k$ epsilons equal 
to $1$ within $2, 4, \ldots , 2n$. This gives us exactly the sum in 
the numerator.

Now one can analyze this explicit formula in the manner of \cite{fe}. 
Asymptotics of every summand in the numerator divided by denominator 
may be found by Stirling's formula. In particular, one observes that for 
"central" binomial coefficients (corresponding to $k=t-1, t+1$ for $n=2t$ 
and $k=t,t+1$ for $n=2t+1$) this asymptotics is $n^{-1/2}$. Now, one can 
expect that the linear factor $\mid 2k - n \mid$ will change asymptotics
to that of $\sqrt n$. This phenomenon is well known is random walks
theory \cite{fe} and it also occurs in our situation.

{\bf Lemma 4}. The average absolute degree of two normal random
hyperbolic binary $n$-forms, for large $n$, is asymptotically equivalent to 
$\sqrt \frac{n}{\pi}$.

We omit details of the proof because it involves some tedious but quite 
standard transformations of this combinatorial sum accompanied by a simple
application of saddle-point asymptotic estimation. The most remarkable
fact here is that this asymptotics turns out to be universal and 
enters into formulation of general result for two variables.

The latter is obtained in a similar way. We only have to take into
account summation with weights equal to probabilities of $A_{rs}$. 
Averages $\tilde d_{rs}$ over $A_{rs}$ are computed using the same 
combinatorics as in Lemma 3. 

{\bf Lemma 5}. For binary $n$-forms with $r$ and $s$ real zeroes, respectively,
the average value of alternances is

\[
\sum_{k=0}^{\frac{r+s}{2}-\frac{\mid r-s\mid}{2}} 
\frac{(C_{k+\mid \frac{r-s}{2} \mid}^{\frac{r+s}{2}})
(C_k^{\frac{r+s}{2}})
\mid 2k + \mid \frac{r-s}{2} \mid - \frac{r+s}{2}\mid}
{C_{\frac{r+s}{2}}^{r+s}} \\.
\]

It is easy to check that order of growth ot these averages is always 
not bigger than that of the average from Lemma 3. Remembering that the
average absolute topological degree is the weighted sum of these averages
with positive "probabilistic" (i.e. summing up to 1) coefficients, we
conclude that the ultimate order of growth will be that of Lemma 3, which
yields the main result of this note.

{\bf Theorem}. The average absolute topological degree of the pair of 
normal random $n$-polynomials, for large $n$, grows as $\sqrt n$.

\section{Concluding remarks}

We would like to indicate that the results above have some immediate 
applications and generalizations. For example, exactly in the same manner 
one can compute the average topological degree of a pair of quasi-homogeneous 
polynomials \cite{avg} and then determine its order of growth for large
values of the weighted algebraic degree. Similar results may be obtained
for random harmonic polynomials and random entire functions.

Also, considering the aforementioned average absolute gradient degree of 
a random $n$-polynomial of two variables, we obtain the growth of $\sqrt{n-1}$. Now, using 
the well known relation 
between gradient degree and Euler characteristic of level surfaces and, 
in particular, formulas from \cite{gkh}, one may apply this result for
finding asymptotics of Euler characteristics of random algebraic surfaces.
Description of underlying topological results and precise statements about 
average Euler characteristic are left for future publications.

Finally, one could try to generalize the theorem to higher dimensions.
Following the lines described at the end of Section 3, it is possible to
guess certain 
higher-dimensional results. We would like to finish this note by the 
formulation of one plausible general hypothesis. Consider the orthogonally
invariant distribution of coefficients of a random $n$-polynomial of
$m$-variables introduced by M.Shub and S.Smale \cite{ssm}. Denote by 
by $g(n,m)$ its average absolute gradient degree.

{\bf Hypothesis}. For fixed $m$ and large $n$, $g(n,m)$ grows as
$(n-1)^{m/4}$.

\end{document}